\documentclass[conference]{IEEEtran}
\IEEEoverridecommandlockouts
\usepackage{cite}
\usepackage{amsmath,amssymb,amsfonts}
\usepackage{algorithmic}
\usepackage{graphicx}
\usepackage{textcomp}
\usepackage{xcolor}
\usepackage{booktabs}
\usepackage{comment}
\usepackage{ragged2e}
\usepackage[utf8]{inputenc}
\usepackage{enumitem}
\usepackage{float}
\usepackage{makecell}
\usepackage{multirow}
\def\BibTeX{{\rm B\kern-.05em{\sc i\kern-.025em b}\kern-.08em
    T\kern-.1667em\lower.7ex\hbox{E}\kern-.125emX}}
\begin{document}


\title{Human-Centered AI Transformation: Exploring Behavioral Dynamics in Software Engineering}

\author{\IEEEauthorblockN{Theocharis Tavantzis}
\IEEEauthorblockA{\textit{University of Gothenburg}\\
Gothenburg, Sweden \\
gustavth@student.gu.se}
\and
\IEEEauthorblockN{Robert Feldt}
\IEEEauthorblockA{\textit{Chalmers University of Technology} \\
Gothenburg, Sweden \\
robert.feldt@chalmers.se}
}

\maketitle

\begin{abstract}
As Artificial Intelligence (AI) becomes integral to software development, understanding the social and cooperative dynamics that affect AI-driven organizational change is important. Yet, despite AI's rapid progress and influence, the human and cooperative facets of these shifts in software organizations remain relatively less explored. This study uses Behavioral Software Engineering (BSE) as a lens to examine these often-overlooked dimensions of AI transformation. Through a qualitative approach involving ten semi-structured interviews across four organizations that are undergoing AI transformations, we performed a thematic analysis that revealed numerous sub-themes linked to twelve BSE concepts across individual, group, and organizational levels.
Since the organizations are at an early stage of transformation we found more emphasis on the individual level.

Our findings further reveal six key challenges tied to these BSE aspects that the organizations face during their AI transformation. 
Aligned with change management literature, we emphasize that effective communication, proactive leadership, and resistance management are essential for successful AI integration. However, we also identify ethical considerations as critical in the AI context—an area largely overlooked in previous research.
Furthermore, a narrative analysis illustrates how different roles within an organization experience the AI transition in unique ways. These insights underscore that AI transformation extends beyond technical solutions; it requires a thoughtful approach that balances technological and human factors.
\end{abstract}

\begin{IEEEkeywords}
Human Aspects, Organizational Change, Artificial Intelligence, AI Transformation, Behavioral Software Engineering
\end{IEEEkeywords}

\section{Introduction}

With the rapid development and impact of 
Artificial Intelligence (AI), understanding the human factors surrounding it has become critical for organizations \cite{fenwick2024critical}. AI’s integration into Software Engineering (SE) is transformative, offering automation, improved productivity, and unlocking new avenues for innovation\cite{ozkaya2023nextFrontier, fan2023llmforSE,yoon2024intent}. 
Furthermore, AI technologies are more than tools for improvement; they are often central components in the design and function of modern software systems. However, SE literature has historically prioritized technological and process-oriented issues over human and cooperative aspects \cite{perry_people_process_improvement_1994, capretz_humanSE_2014,storey2020software}.
This focus risks repeating itself as AI reshapes the field, potentially overlooking the critical human dynamics necessary for effective transformation.
We suggest that frameworks and methods rooted in socio-technical perspectives~\cite{storey2020software,hoda2021socio} or those incorporating psychological, cognitive, behavioral, and social theories and concepts, e.g. Behavioral Software Engineering (BSE)~\cite{lenberg_bse_2014,lenberg_bse_slr_2015}, can provide valuable insights into the complex, human-centered challenges of AI-driven transformations in software organizations.

As organizations undergo change driven by AI, human aspects are becoming vital for the success of these transitions~\cite{fenwick2024critical}. 
Traditional change management models, such as Kotter's~\cite{kotter_leading_change_2012}, often overlook the behavioral and psychological factors that shape how employees adapt to change. While recent research has explored human aspects of successful organizational change~\cite{kotter_heart_2012}, including in software organizations~\cite{lenberg_human_factors_2015, lenberg_organizational_change_2017}, these studies predate any substantial impact of AI. Moreover, a search for studies on ``AI transformation'' combined with behavioral or human factors yields few results, with no direct references to Behavioral Software Engineering (BSE). Peretz-Andersson and Torkar’s systematic mapping study~\cite{peretz_andersson_empirical_2022} confirms this gap, finding psychology-related studies underrepresented in current AI transformation research.

This study empirically investigates the human and behavioral dynamics underpinning AI-driven transformations in organizations. Given the nascent state of research in this area, we posit that insights from industry practitioners experiencing these shifts firsthand are both timely and invaluable, offering a more nuanced perspective. To capture a broad spectrum of experiences, we interviewed practitioners across multiple organizations and roles, using semi-structured interviews. 

We framed our analysis through a Behavioral Software Engineering (BSE)~\cite{lenberg_bse_slr_2015} lens, leveraging it to structure and enhance our understanding of the data.
Given the ongoing nature of AI transformation, we concentrated on challenges related to BSE concepts rather than focusing on or proposing specific solutions. Anticipating that individuals’ perspectives might differ based on their roles and experiences, we paid particular attention to these variations.

The structure of the rest of this paper is organized as follows: Section II
provides an overview of the related work, while Section III explains the research methodology that was followed. Section IV presents the results of the study, addressing the research questions. Then, Section V discusses the study’s findings, and, finally, Section VI outlines the conclusions and potential directions for future work.

\section{Related Work}

\subsection{Behavioral Software Engineering}

Human factors in Software Engineering (SE) have been recognized as essential since the field's early days \cite{weinberg_psychology_1971}. However, despite early acknowledgment, much of SE research and practice has concentrated on technical processes, often overlooking the social and psychological dimensions of the SE life cycle \cite{perry_people_process_improvement_1994, ferreira_spi_2011, capretz_humanSE_2014}. In response, Lenberg et al. introduced the concept of Behavioral Software Engineering (BSE) in 2014 to focus explicitly on these human aspects \cite{lenberg_bse_2014}.

Building on this foundation, Lenberg et al. conducted a systematic literature review in 2015 to map BSE research. Their findings showed growing interest in human-centered SE topics but also revealed significant research gaps. Notably, 42 out of 55 identified BSE concepts had fewer than 10 publications, indicating substantial room for further study \cite{lenberg_bse_slr_2015}. Since then, researchers have examined various BSE areas, including emotions \cite{graziotin_unhappiness_2017}, personality \cite{felipe_personality_2023}, and attitudes and norms \cite{gren_attitudes_norms_2018}. Among these, Organizational Change has gained particular interest \cite{lenberg_organizational_change_2017, understanding_org_change_2023, klotins2022continuous}—a trend likely to intensify as technology continues to evolve rapidly.

\subsection{Organizational Change}

Organizations typically undergo transformations to align with market demands, while maintaining productivity and quality \cite{lima_organizational_2023}. The incentive for such a change often stems from internal factors, like productivity improvement, or external factors, such as emerging technologies \cite{beer_walton_organization_1987}.

Although existing research emphasizes the importance of considering human factors in organizational change, this focus often is not reflected in practice \cite{roy_human_1998, ferreira_spi_2011}. In their study, Roy et al. \cite{roy_human_1998}, analyzed four case studies and conducted a survey to determine which human factors are considered in Business Process Re-engineering (BPR) initiatives. Their findings indicated that while BPR projects achieved their objectives, the human aspects were frequently overlooked. However, the research did not provide insights on the impact of these changes on employees' satisfaction and well-being, factors that are critical for individuals' performance and productivity, particularly in software development \cite{graziotin_unhappiness_2017}.

Serour and Henderson-Sellers \cite{serour_resistance_2005} explored the impact of considering human factors during an organization’s transition to Object Technology, with a primary focus on employee resistance to change. The findings demonstrated that neglecting human factors led to the failure of the initial change effort, whereas the successful second attempt was attributed to effective planning and management of human elements, resulting in greater acceptance of the change.

Later on, Lenberg et al. conducted a qualitative study, highlighting that successful transformations within the field of Software Engineering require a comprehensive understanding of cognitive, behavioral, and social factors  \cite{lenberg_human_factors_2015}. In a subsequent study, Lenberg et al. \cite{lenberg_organizational_change_2017} investigated the link between positive attitudes towards the transformation and the success of the change initiatives. In particular, they showed that the \textit{knowledge, need for change, and participation to change} are key enablers to the \textit{openness to change}, with the latter two being significant for \textit{readiness for change}. Based on the findings, they made the assumption that a positive attitude towards Organizational Change is crucial for its success.

Similarly, Errida and Lotfi conducted a mixed-method study \cite{Errida2021DescriptiveOrgModel}, combining a systematic literature review with action research, to identify success factors in change management projects. Specifically, they reviewed 37 existing models. The authors ended up with 12 categories that are, primarily, focusing on the behavioral aspects of change management. The findings of the action research highlighted that the leadership of the change manager, the effective and constant communication, stakeholders' engagement, and the motivation of employees and change agents were the most critical factors for a successful change management project.

\subsection{Artificial Intelligence Transformation}

Artificial Intelligence (AI), as a game-changer, has gained increasing attention from both academia and industry. Haefner et al. introduced a framework for implementing and scaling AI \cite{haefner_scalingAI_2023}. In this context, they discussed the socio-technical dimension of such a transformation. On the social side of the change, organizations need to design a long-term strategy for integrating AI, shifting their focus from vertical capabilities to horizontal capabilities, which facilitate scalability across different business units. In this way, AI integration is ensured, enabling the organizations to fully leverage the benefits of such a technology.

Oldemeyer et al. discussed the challenges of the AI integration by small and medium enterprises \cite{oldemeyer_ai_smes_2024}. They identified 27 different challenges via their systematic literature \textit{political, economic, social, technological, environment, and legal}. The social category contained the challenges of Management, Acceptance, AI corporate strategy, and Trust/Ethic. Although the authors were discussing the social challenges, they did not delve into the psychological aspects of those challenges.

On a similar note, Schwaeke et al. \cite{schwaeke_ai_adoption_2024} conducted a systematic literature review to present the status quo of AI adoption by small and medium enterprises by introducing an updated version of an existing Technology, Organization, and Environment (TOE) framework. In this way, the authors aimed to illustrate the complex and multifaceted project of AI integration, as it can be affected by a range of factors and variables. The proposed TOE framework, as a more high-level depiction, discusses social aspects, like resistance, but does not delve more into social and psychological dimensions.

\subsection{Human Aspects in AI Transformation}

Despite the rapid rise of AI, significant research gaps persist in this interdisciplinary field, particularly concerning human factors. Peretz-Andersson and Torkar \cite{peretz_andersson_empirical_2022} mapped the existing body of AI transformation research, identifying only 4 psychology-related studies among the 52 reviewed. However, these few studies do not necessarily address the behavioral aspects critical to AI-driven change. In particular, they discuss the user perspective \cite{feng2019understanding} or focus on the importance of soft skill in adopting Big Data technologies \cite{caputo2019innovating}, among others. Consequently, the authors highlight a clear gap, calling for future research into the human dimensions of AI transformation.

A systematic review by Dwivedi et al. \cite{dwivedi_AI_2021} adds to this perspective by examining the social and ethical challenges posed by AI. They discuss potential job displacement due to AI-driven automation and ethical issues, such as algorithmic bias and opaque decision-making, which align with BSE’s focus on social impact at various levels. The authors conclude that interdisciplinary research—incorporating social sciences and psychology—is essential to fully understand AI’s broader societal implications, echoing Peretz-Andersson and Torkar’s findings \cite{peretz_andersson_empirical_2022}.

Singh \cite{satpreet_leadershipAI_2023} approaches AI transformation through an organizational lens, analyzing how leadership models should adapt to AI-driven change. He argues that leaders should foster a human-centered approach, emphasizing continuous communication, learning, feedback, and ethical practices. For software developing organizations, this study thus underscores the need for BSE research on AI transformation, not only at the individual, but also at the group and organizational levels.

\section{Research Method}

\begin{figure}[htbp]
\centerline{\includegraphics[scale=0.42]{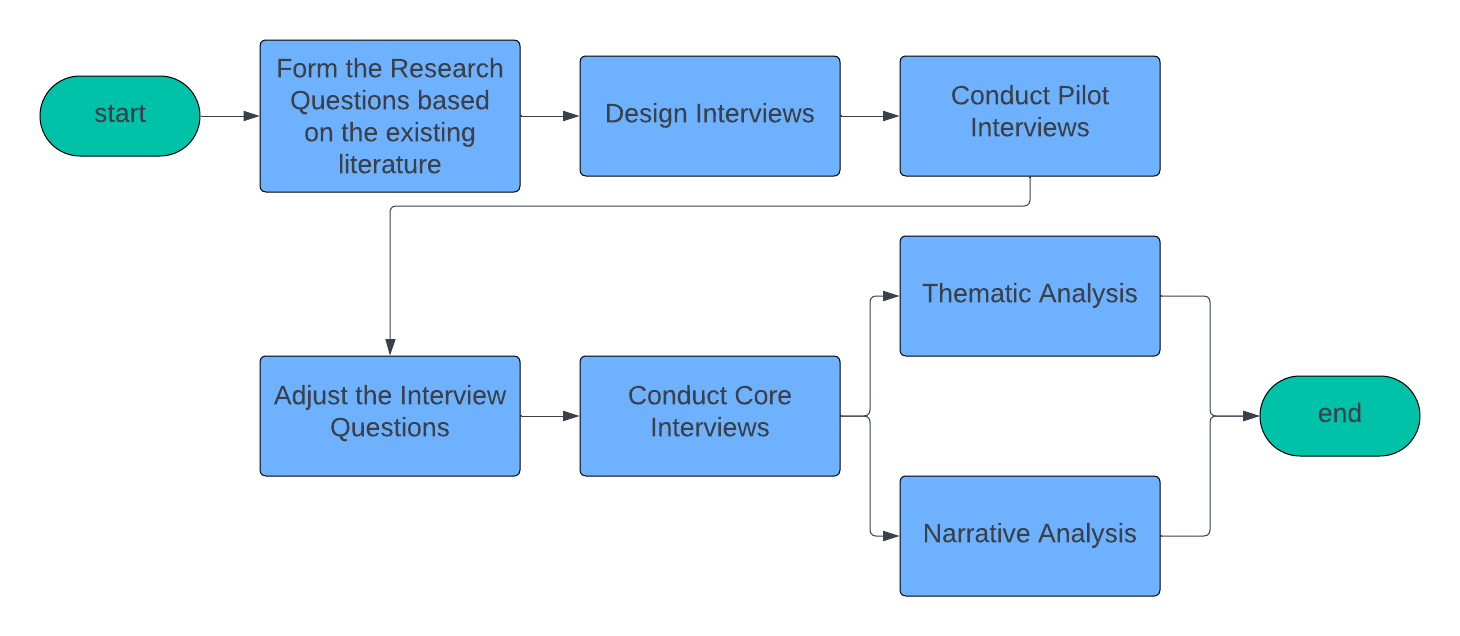}}
\caption{Research Method Overview}
\label{fig:research_method}
\end{figure}

The primary objective of this study is to examine how behavioral and social dynamics are influencing the AI-driven changes of organizations where software development or software solutions is a key element of their ''offering''. Since little is yet known about these transformation processes we conducted an exploratory study to help shape future research and guide the development and evaluation of practical solutions. Thus, we conducted semi-structured interviews (see Figure \ref{fig:research_method}) across a diverse range of Swedish organizations. These organizations are in the initial stages of AI integration, as only a few institutions have reached high levels of maturity in this area. Interviewees represented various roles within these organizations, providing a broad and nuanced view of the behavioral factors influencing AI-driven organizational change. This qualitative approach can offer valuable insights into the unique challenges and adaptations SE professionals face as they navigate AI’s impact on their work environments.

\subsection{Research Questions}

This study aims to clarify the role of BSE concepts in AI transformations by identifying which behavioral and social factors are essential for successful AI adoption, as well as the primary challenges involved. To guide our investigation, we developed three main research questions:\\

\textbf{- RQ1:} What are the \textbf{key BSE concepts} that influence AI-driven software organizational change?

\textbf{- RQ2:} What challenges are associated with these key BSE concepts, as identified in \textit{RQ1}?

\textbf{- RQ3:} How do \textbf{different organizational roles} experience and navigate the AI transformation process?

Through these questions, the study seeks to pinpoint the behavioral dimensions critical for effective AI integration in software-developing or software-dependent organizations.

\subsection{Data Collection and Interviewees}

The data collection process included two phases: pilot interviews and core interviews. The pilot interviews had two main objectives: first, to identify key concepts relevant to the research questions, and second, to refine the interview format based on initial insights. We used semi-structured interviews, balancing structured questions with the flexibility to explore emerging topics in depth.

We conducted ten interviews with participants from four organizations in the Gothenburg area, each actively engaged in AI transformation and either developing their own software products or being heavily dependent on software for their core activities. Table \ref{tab:organizations_description} provides more details about these organizations, which were selected through existing or prior university collaborations. These connections facilitated access to key personnel. We used convenience sampling, specifying that each organization include at least one interviewee involved in AI project development and one involved at a higher level, related to the AI transformation strategy.

\begin{table}[h!]
    \centering
    \scalebox{0.73}{
    \begin{tabular}{|c|c|c|c|c|}
        \hline
        \textbf{Org. ID} & \textbf{Part. ID} & \textbf{Company Size} & \textbf{Type of Organization} & \textbf{Domain of Organization} \\ 
        \hline
        O1 & P1, P2 & 6,700 employees & software-dependent & public sector\\
        \hline
        O2 & P3, P6, P10 & 10,000 employees & software-developing & automotive\\
        \hline
        O3 & P4, P5 & 22,000 employees & software-developing & aerospace\\
        \hline
        O4 & P7, P8, P9 & 12,000 employees & software-developing & medtech\\
        \hline
    \end{tabular}
}
\caption{descriptive information of the organizations}
\label{tab:organizations_description}
\end{table}

Interviews took place either online (five out of ten) or on-site (five out of ten). The analysis was performed on all ten interviews, as even the pilot one disclosed useful insights. Participants represented a variety of roles and departments, capturing a broad perspective on organizational experiences with AI adoption. This diverse sample enabled a comprehensive view of behavioral dynamics across different areas within each organization.
Table \ref{tab:participants_description} offers detailed participant information.

\begin{table}[h!]
    \centering
    \scalebox{0.65}{
    \begin{tabular}{|c|c|c|c|c|}
        \hline
        \textbf{Part. ID} & \textbf{Role} & \textbf{Work Exp.} & \textbf{Years in Current Role} & \textbf{Experience with AI} \\ 
        \hline
        P1 & Research Coordinator & 11 & 1 year & High \\
        \hline
        P2 & IT Support Section Manager & 15 & Less than 1 year & High \\
        \hline
        P3 & \makecell{General Assistant for AI\\Team Operations} & 2 & Less than 1 year & High \\
        \hline
        P4 & System Engineer & 12 & 10 & Low \\
        \hline
        P5 & System Engineer & 17 & 1 & Low \\
        \hline
        P6 & Software Developer & 13 & 2 & High  \\
        \hline
        P7 & \makecell{Head of Global Leadership\\and Executive Development} & 21 & Less than 1 year & High \\
        \hline
        P8 & \makecell{Organizational Development\\Manager} & 8 & 2 & High \\
        \hline
        P9 & \makecell{Director of HR and\\Business Analytics} & 21 & 4 & Low\\
        \hline
        P10 & Software Developer & 12 & 5 & High  \\
        \hline
    \end{tabular}
}
\caption{descriptive information of the participants}
\label{tab:participants_description}
\end{table}

\subsection{Data Analysis}

The study employed \textit{thematic analysis} to analyze the qualitative data from the interviews, following the method outlined by Braun and Clarke \cite{braun2006thematicAnalysis}. This approach enabled us to identify, analyze, and report patterns across the interview transcripts. To enhance reliability, two researchers independently identified codes and themes of one interview, resolving any discrepancies through discussion. Due to time limitations only one researcher then analyzed the rest of the interviews.

To analyze the qualitative data for RQ1, we relied our theme definition in established concepts from the existing BSE literature, ensuring consistency with prior research. Then, the sub-themes were delineated using Braun and Clarke’s methodology \cite{braun2006thematicAnalysis}. Conversely, for RQ2 both the theme and sub-theme definition were grounded solely on the systematic process for thematic analysis, enabling themes and sub-themes to emerge naturally from the analysis itself.

To complement the thematic analysis, achieve methodological triangulation, and address the final research question -- how different roles experience AI transformation -- we applied \textit{narrative analysis}. Narrative analysis examines how people construct meaning through the stories they share \cite{murray2008NarrativeAnalysis}. According to Murray, narrative analysis unfolds in two phases, which we also followed: descriptive and interpretive \cite{murray2008NarrativeAnalysis}. In the descriptive phase, researchers identify the narratives’ beginning, middle, and end, along with their context. In the interpretive phase, they connect these narratives to the study’s context, analyzing them to derive insights.

In this study, narrative analysis helped us uncover how individuals’ roles influence their experiences of AI-driven organizational change, highlighting connections between their perspectives and their positions within the organization. Following Murray's methodology \cite{murray2008NarrativeAnalysis}, we categorized narratives as regressive, stable, progressive. In the context of the study, a regressive narrative indicated that individuals perceived their roles as being negatively impacted by AI transformation, focusing primarily on adverse experiences such as stress or job insecurity. A progressive narrative, on the other hand, described AI as a catalyst for positive change and personal or professional growth. Finally, a stable narrative reflected a perspective where AI was seen as an additional tool that did not significantly alter the individual's role.

\subsection{Threats to Validity}

As with any research, this study faces several threats to validity.

\textbf{\textit{Construct Validity:}} A potential threat to construct validity is that the interview questions were developed by the research team. This could introduce unintended biases in question framing. Additionally, thematic analysis depends on the researchers’ interpretations, which may also introduce bias when identifying data patterns. To mitigate these risks, we employed a systematic approach to constructing the interview questions, and both authors participated in key parts of data analysis, holding multiple discussions to minimize interpretive bias. Another concern is social desirability bias, where respondents might provide socially acceptable answers instead of their genuine thoughts. This issue also affects internal validity, as it can compromise the study’s ability to capture authentic participant perspectives. To address this, we assured participants of response confidentiality, encouraging honest feedback.

\textbf{\textit{External Validity:}} The study’s findings are limited by the relatively small number of participating organizations, all within a narrow domain. While we included one organization which did not themselves develop software but was dependent on software for their core activities, the small total number may still affect the generalizability of results, as organizations with different cultures or operational domains might experience AI integration differently. To counter this, we included organizations from diverse domains (medtech, aerospace, and automotive) and selected employees with a range of roles and backgrounds to broaden the scope and insights into AI-driven organizational change. 

\textbf{\textit{Internal Validity:}} Internal validity may be affected by sample size and selection bias. Only a few companies have yet gotten significant experience with AI transformation, which could introduce selection bias and limit the study’s representation of the full complexity involved.

\section{Results}

\subsection{RQ1: BSE concepts that influence the AI Organizational Change}

Below we go through the concepts and their associated sub-themes that surfaced during the interviews. For each sub-themes we exemplify with a quote from one of the related interviews.

Our analysis resulted in twelve themes related to BSE concepts, most of which have been recognized within the BSE literature. Thus, consistency with the SE literature is maintained. Each theme includes multiple sub-themes, which collectively contribute to the overall understanding of BSE-driven AI transformation. The themes and sub-themes are ordered alphabetically, as no weight strategy was defined. The findings are presented in three different Tables, grouping the themes and sub-themes based on the three different levels of analysis, namely individual (Table \ref{tab:BSE_individual}), group (Table \ref{tab:BSE_group}), and organizational (Table \ref{tab:BSE_organizational}).

\begin{table}[h!]
    \centering
    \scalebox{0.65}{
    \begin{tabular}{|c|c|c|}
        \hline
        \textbf{Theme} & \textbf{Sub-theme} & \textbf{Summary} \\ 
        \hline
        \multirow{5}{*}{Attitudes} 
        & Contribute and Embrace the change & \multirow{5}{6cm}{The sub-themes related to \textit{Attitudes} refer to the individuals' attitudes towards AI and how those attitudes evolved through time.} \\
        \cline{2-2}
        & Disapproval for AI use & \\
        \cline{2-2}
        & Initial skepticism towards AI & \\
        \cline{2-2}
        & Progressive Attitude shift towards AI & \\
        \cline{2-2}
        & Positivism for AI growth & \\
        \cline{2-2}
        & Skepticism towards Tech Industry & \\
        \hline
        \multirow{3}{*}{Cognitive} 
        & Confusion regarding AI & \multirow{3}{6cm}{The sub-themes related to \textit{Cognitive} highlight the misunderstandings that people face regarding AI.} \\
        \cline{2-2}
        & \makecell{Concerns regarding job\\displacement and roles evolution} & \\
        \cline{2-2}
        & Misconception of AI solution's suitability & \\
        \hline
        \multirow{3}{*}{Creativity} 
        & Brainstorming with AI & \multirow{3}{6cm}{The sub-themes related to \textit{Creativity} discuss how AI can inspire new ideas but may also stifle creativity, if overused.} \\
        \cline{2-2}
        & Creates new possibilities & \\
        \cline{2-2}
        & \makecell{Decreased Creativity due to\\over-dependence on AI} & \\
        \hline
        \multirow{14}{*}{Emotions} 
        & Disappointment due to low AI performance  & \multirow{14}{6cm}{The sub-themes related to \textit{Emotions} summarize the multifaceted feelings that individuals deal with during the AI-driven Organizational Change.} \\
        \cline{2-2}
        & Excitement due to potential of AI & \\
        \cline{2-2}
        & \makecell{Excitement due to impact\\of AI on workflow} & \\
        \cline{2-2}
        & Fear of AI’s growth & \\
        \cline{2-2}
        & Fear due to lack of understanding & \\
        \cline{2-2}
        & Frustration due to existing limitations & \\
        \cline{2-2}
        & Frustration due to resistance to change & \\
        \cline{2-2}
        & Frustration due to specific AI use & \\
        \cline{2-2}
        & Guilt due to AI use & \\
        \cline{2-2}
        & \makecell{Overwhelmed Feelings due to\\pace of change} & \\
        \cline{2-2}
        & \makecell{Sadness due to a specific type of AI use} & \\
        \cline{2-2}
        & Surprise due to quick AI adoption & \\
        \cline{2-2}
        & Worry about specific AI use & \\
        \cline{2-2}
        & Worry about market disparities & \\
        \hline
        \multirow{3}{*}{Motivation} 
        & \multirow{3}{*}{Automate monotonous tasks} & \multirow{2}{6cm}{The sub-theme related to \textit{Motivation} discusses the factors that lead to AI leveraging by individuals.} \\
        & & \\
        & & \\
        \hline
        \multirow{4}{*}{Personality} 
        & \multirow{2}{*}{Openness to change} & \multirow{4}{6cm}{The sub-themes related to \textit{Personality} refer to personality traits that influence individuals' approach towards AI.} \\
        & & \\
        \cline{2-2}
        & \multirow{2}{*}{Curiosity} & \\
        & & \\
        \hline
        \multirow{4}{*}{Stress} 
        & \multirow{2}{*}{Multiple concurrent projects} & \multirow{4}{6cm}{The sub-themes related to \textit{Stress} reflect the pressures that individuals face from AI-driven projects.} \\
        & & \\
        \cline{2-2}
        & \multirow{2}{*}{Frequent plan adjustments} & \\
        & & \\
         \hline
    \end{tabular}
    }
    \caption{identified themes and sub-themes in individual level}
    \label{tab:BSE_individual}
\end{table}

Since most organizations are in the early stages of AI transformation, participants provided more insights into AI's impact on individuals than on group workflows or organizational structures. Consequently, the individual level comprises seven of the twelve identified themes (Table \ref{tab:BSE_individual}). Specifically:

\textbf{\textit{Attitudes:}} Participants revealed diverse attitudes toward AI during its early adoption phase. Initially, skepticism was common; half of the participants expressed doubts. For instance, P4 noted, \textit{"I was really like, okay, I'm not sure. Let's see what is it exactly? What can it be used for?"}. However, as participants engaged more with the technology, many shifted from doubt to acceptance. P3 explained,  \textit{"Then, we had some really good lecturers who encouraged us to use it... And it changed my attitude towards AI in a positive way. So, I started using it as a trampoline to help bounce ideas back and forth and help my thinking process."}. 

Despite this general trend, some participants retained negative perceptions. P8 mentioned,  \textit{"there's like a bit of stigma about people relying on it for their roles"}, while P9 expressed a general skepticism towards tech industry by saying \textit{"I do it so that there's no kind of connection to the company simply because I think the tech industry cannot be trusted"}. Conversely, other participants were optimistic from the outset, viewing AI as a catalyst for growth. P6, for example, remarked,  \textit{"But personally, I would think that maybe it would make few redundant jobs obsolete, but then it would also give rise to new jobs. For example, until few years back, nobody was aware of the role called prompt engineer."}. Similarly, some respondents, like P7, stressed the inevitability of progress:  \textit{"But I don't think you can stop progress, it's going to happen. So we've got to just embrace it and move on to the next thing"}.

\textbf{\textit{Cognitive:}} Three sub-themes that emerged from our analysis can be grouped as cognitive aspects. Many interviewees expressed that there is much confusion about how AI functions and how it might complement human work. P7 observed, \textit{"I think there is confusion and how can AI help me is usually the question. So when we talk about our vision... not so many people are there yet."}. Similarly, P4 remarked, \textit{"...what I always find quite funny these days is the case that you say 'oh yeah this is AI-driven' and then sometimes I'm like 'you know, this is just an algorithm like the ones we had before'"}. P6 acknowledged that there are misconceptions of when AI constitutes a suitable solution, noting, \textit{"Unfortunately, some people think that everything can be done in AI."}. Finally, P4 voiced concerns about \textit{"replacing all service centers with chatbots."}

\textbf{\textit{Creativity:}} Several participants conveyed the view that AI can enhance creativity, helping them to think outside the box. P2 shared, \textit{"For me, it's more like it boosts my creativity... it helps me to do things I hadn't thought were possible before."} However, P5 warned that over-reliance on AI could dampen innovation: \textit{"...like any other tool, it can go both ways. Either you rely too much on it and lose creativity or you find a creative way to use it."}

\textbf{\textit{Emotions:}} The emotional impact of AI transformation varied widely, from excitement to frustration. Participants expressed enthusiasm about AI’s potential and its positive impact on daily tasks; for example, P2 noted, \textit{"I really don’t miss writing meeting notes."}. However, fear was also prevalent, often tied to a lack of understanding of AI’s capabilities or its possible consequences. P7 described it as a \textit{"fear of the unknown,"} while P1 worried about job displacement and funding cuts, saying, \textit{"I think there's a worry like what's going to happen here, there's research funding being withdrawn..."}. P5 expressed concern over market competition, adding, \textit{"I don't see any way that a smaller company can compete with that."}

Feelings of guilt also surfaced, as P7 mentioned, \textit{"Somebody mentioned the guilt of using it (AI)... I should be doing this myself"}. Others, like P3, felt sadness over the \textit{"the replacement of ''artistic'' tasks,"}, while P9 expressed their disappointment as \textit{"AI's supposed to be so much smarter than just me google searching, which seems not to be the case."}.

Frustration was common as well, often due to AI’s limitations or misuse. P8 cited instances where  \textit{"when it's misused or incorrectly applied or people just adopt it and come across as inauthentic."}. P7 highlighted internal resistance to AI adoption, saying, \textit{"You've always got the early adopters and then the 'I’m never gonna do this'. So that's always, a frustration."}. Finally, P7 shared a sense of being overwhelmed by the rapid pace of change, remarking, \textit{"He was saying he spends three hours per day keeping up to date on the changes. Okay, that's a lot, right? So yeah, it's almost impossible."}

\textbf{\textit{Motivation:}} For most interviewees, the motivation to adopt AI was often tied to personality traits like curiosity. However, few participants also cited a desire to automate repetitive tasks to boost productivity. As P6 explained, \textit{"if there is a tool that could automatically write test cases in Python, that would enable app developers to focus more on the application quality."}

\textbf{\textit{Personality:}} During interviews it was clear that the personality traits and characteristics of the participants played a significant role in how they approached and also integrated AI into their workflows. Curiosity was common, driving exploration and experimentation with new tools. As P1 shared, \textit{"...when I'm getting a new license, I want to dive in and explore."}. Openness to change also influenced individuals’ interactions with and view of AI tools. P5 described their approach as \textit{"conservative, following the operation context of the business"}, while P6 was more adaptive, stating, \textit{"Personally, I am very open to adopting new technologies, especially AI..."}.

\textbf{\textit{Stress:}} Stress emerged as a recurring theme in the context of AI-driven organizational change. Multiple concurrent projects added to this stress, as P1 noted,  (\textit{"...it can become stressful, because then I have to move all of those projects forward."}). Additionally, P6 described how frequent plan adjustments within their organization created tension, saying, \textit{"Initially, we started with an outdated model...then they came up with a newer model... So, all these things have messed up the timelines."}).

As the AI transformation remains in its early stages for most involved organizations, the group-level impact seems to, so far, has been limited. Our analysis identified only two main themes, see Table \ref{tab:BSE_group}.

\begin{table}[h!]
    \centering
    \scalebox{0.65}{
    \begin{tabular}{|c|c|c|}
        \hline
        \textbf{Theme} & \textbf{Sub-theme} & \textbf{Summary} \\ 
        \hline
        \multirow{4}{*}{Group Dynamics} 
        & AI-driven thinking & \multirow{4}{6cm}{The \textit{Group Dynamics} theme describes how AI has influenced teams collaboration and mindset.} \\
        \cline{2-2}
        & Driving Agility & \\
        \cline{2-2}
        & Cross-functional teams & \\
        \cline{2-2}
        & Engaging collaboration & \\
        \cline{2-2}
        & \makecell{Substituting presence\\and collaboration} & \\
        \hline
        \multirow{3}{*}{Politics} 
        & \multirow{2}{*}{Differing approaches within groups} & \multirow{3}{6cm}{The \textit{Politics} theme discusses the varying approaches and perspectives regarding AI within groups, and how these can drive or challenge teamwork.} \\
        & & \\
        \cline{2-2}
        & Generational differences & \\
        \hline
    \end{tabular}
    }
    \caption{identified themes and sub-themes in group Level}
    \label{tab:BSE_group}
\end{table}

\textbf{\textit{Group Dynamics:}} The analysis revealed that AI has already begun to reshape how groups work and collaborate, moving away from conventional practices. P6 described this shift in mindset as more AI-driven: \textit{"...we have started to think in an AI way; earlier we were thinking of a traditional approach, maybe using some Python script... but now we are thinking if this problem can be solved by AI or not."}) This transformation has also underscored the need for agility, given the rapid pace of AI advancements. In this context, P3 observed,  \textit{"...the nature of working with AI in our group leads to be more agile, because things are happening so fast in AI environment... So, I mean, it's definitely like a heightened need to be quick on our feet and be agile."} 

In addition, four participants emphasized that AI’s multifaceted nature benefits from cross-functional team collaboration. P1 highlighted the importance of networking across different backgrounds, saying,  \textit{"I think it is crucial to network and communicate with different backgrounds, because then you can make things happen much easier and faster."}, while P3 referred to AI as a \textit{"very cross-functional topic"}. This cross-functional approach positively influenced group dynamics as well, with P2 noting,  \textit{"...and that made the dynamic more engaging and more fun, because we can try out new technology and be part of building a product for ourselves."} 

However, the impact of AI on group dynamics also presented challenges. P8 described instances where AI solutions were \textit{"substituting human presence and hindering collaboration during meetings"}.

\textbf{\textit{Politics:}} Although typically discussed at the organizational level, some sub-themes related to politics emerged within or between groups, so we report them here at the group level of analysis.

Four participants highlighted differences in approach among roles within a group. P5 contrasted developers and product managers, noting that developers are \textit{"more early-adopters and experiment more"}, while product managers often stick to familiar tools:  \textit{"I have these five tools, I apply those five tools to my projects, and they're tools of Microsoft's office."}. 
Additionally, P2 observed that generational differences influenced AI adoption within groups, explaining,  \textit{"...everyone is 10 or 20 years older than me. And they don't use the models as much as me. So sometimes when I try to describe how we should adapt the department, they don't really relate to what I'm speaking about..."}.

The thematic analysis of the interview transcripts identified three distinct themes and BSE concepts at the organizational level: \textit{Organizational Culture, Organizational Readiness, and Politics}.
These themes, along with their sub-themes, are presented in Table \ref{tab:BSE_organizational}.

\begin{table}[h!]
    \centering
    \scalebox{0.61}{
    \begin{tabular}{|c|c|c|c|c|}
       \hline
        \textbf{Theme} & \textbf{Sub-theme} & \textbf{Summary} \\ 
        \hline
        \multirow{9}{*}{Organizational Culture} 
        & Build internal AI solutions & \multirow{9}{6cm}{The \textit{Organizational Culture} theme refers to the company’s approach to AI adoption, emphasizing on values and practices that have been identified during the AI Transformation journey.} \\
        \cline{2-2}
        & Collaboration with academia & \\
        \cline{2-2}
        & Company’s Continuous Improvement & \\
        \cline{2-2}
        & Conservative organizational culture & \\
        \cline{2-2}
        & Experimentation culture & \\
        \cline{2-2}
        & General framework for AI use & \\
        \cline{2-2}
        & Knowledge-sharing culture & \\
        \cline{2-2}
        & Lack of transparency in AI Strategy & \\
        \cline{2-2}
        & Over-adoption of AI Solutions & \\
        \cline{2-2}
        & \makecell{Support Employee’s\\Continuous Improvement} & \\
        \hline
        \multirow{3}{*}{Organizational Readiness} 
        & \multirow{3}{*}{\makecell{Misalignment between planning and\\latest AI progress}} & \multirow{3}{6cm}{The \textit{Organizational Readiness} theme relates to the company's capability of aligning the organizational strategy with the latest AI advancements.} \\
        & & \\
        & & \\
        \hline
        \multirow{3}{*}{Politics} 
        & \makecell{Divergent Departmental Attitudes\\towards AI integration} & \multirow{3}{6cm}{The theme of \textit{Politics} explores issues which, potentially, can affect the overall organization's strategy and integration process.} \\
        \cline{2-2}
        & External Pressure for AI integration & \\
        \cline{2-2}
        & Internal Pressure for AI integration & \\
        \cline{2-2}
        & Silo work among departments & \\
        \hline
    \end{tabular}
    }
    \caption{identified themes and sub-themes in organizational level}
    \label{tab:BSE_organizational}
\end{table}

\textbf{\textit{Organizational Culture:}} Participants discussed several aspects of organizational culture that influence or is influenced by AI-driven change. All organizations involved provide internal AI solutions and guidelines for using external tools like ChatGPT. For instance, P5 noted,  \textit{"...and there you have setup rules for what you can and cannot share so you can still not place company's private data."}. 

Participants also highlighted varying approaches to business strategy. P4, P5, and P7 described a conservative stance toward change, with P5 remarking,  \textit{"...the industry we're working with is very conservative. We're generally not applying anything that is cutting edge... We actually have a technical standard from the 1940s.")}. In contrast, P3 and P6 outlined an innovation-focused strategy aimed at continuous improvement, with P3 stating,  \textit{"I think that we have very ambitious goals and I think just AI presents so many opportunities like being a high-tech product development company."}. P6, among others, emphasized the value of a knowledge-sharing culture within their organization, mentioning \textit{"forums in \textbf{Org. X} that contribute to advertising AI or sharing knowledge on AI and best practices}. 

Employee development initiatives were widely mentioned. P9 highlighted \textit{"education programs and workshops"} provided for employees, while P6 discussed a specific program that \textit{"teaches the employees how to query ChatGPT using prompt engineering"}. P5 shared their company’s \textit{"philosophy to collaborate with universities for research"}. However, P5 also criticized the trend of overusing AI, stating,  \textit{"If all you have is a hammer and every problem looks like a nail, you are applying AI to everything. And a lot of companies seem to have that approach."}. 

Finally, P8 underscored the confusion caused by insufficient transparency around AI strategies, emphasizing, \textit{"that's where the transparency of our strategy around these topics is really important. And I think that we need to do more on that."}

\textbf{\textit{Organizational Adaptability:}} P6 described a situation where the organization’s planning was not aligned with the latest AI advancements, resulting in project delays. They noted, \textit{"there had been this newer technology when we started the project, and if we had adapted from the beginning, we would have been in a much better position"}.

\textbf{\textit{Politics:}} In the interviews it was clear that organizational politics played a role in AI integration, with departments taking varied approaches. P2 noted, \textit{"when we talk to other departments... they don’t want to use LLM models because they’re scared the products won’t give them the right answer".} Such discrepancies, according to P1 and P2, can lead to project delays. P3 described instances where  \textit{"departments want to work on the same idea, but without collaborating,"}leading to siloed efforts that may drive up costs and later require bridging gaps between teams. 

Additionally, P6 and P8 highlighted internal and external pressures that spurred the AI transformation in their organization. P6 mentioned,  \textit{"internal pressure comes from the employees themselves, as, for example, they can still use external LLMs, even if there are restrictions."} Meanwhile, P8 noted that rapid AI advancements have \textit{"put even more expectation on our shoulders, because we need to be seen by our investors and our stakeholders and all of our groups to be grasping it."}

\subsection{RQ2: Challenges related to the identified BSE concepts}

The thematic analysis of the interview transcripts identified six distinct themes related to the challenges of a BSE-driven AI transformation. The results are presented in Table \ref{tab:challenges} and exemplified below. 

\begin{table}[h!]
    \centering
    \scalebox{0.63}{
    \begin{tabular}{|c|c|c|}
       \hline
        \textbf{Theme} & \textbf{Sub-theme} & \textbf{Summary} \\ 
        \hline
        \multirow{2}{*}{Change Management Strategy} 
        & External communication & \multirow{2}{5.2cm}{This theme refers to the challenges, when it comes to communicating the change.} \\
        \cline{2-2}
        & Internal communication & \\
        \hline
        \multirow{3}{*}{Ethical Concerns} 
        & Competitive gap among the market & \multirow{3}{5.2cm}{The theme of \textit{Ethical Concerns} refers to ethical issues that arise during the AI-driven Organizational Change.} \\
        \cline{2-2}
        & \makecell{Considerate use of external AI\\with company’s data} & \\
        \cline{2-2}
        & Risk aversion due to sensitive data & \\
        \hline
        \multirow{4}{*}{\makecell{Organizational Readiness for\\AI Adoption}}
        & \makecell{Adaptation to global and local\\data legislation} & \multirow{4}{5.2cm}{The \textit{Organizational Readiness for AI Adoption} theme relates to the importance of managing organizational issues, e.g. diverse views and sector-specific complexities, for a smooth AI adoption.} \\
        \cline{2-2}
        & Cross-Functional Collaboration & \\
        \cline{2-2}
        & Ensuring organizational alignment & \\
        \cline{2-2}
        & \makecell{Handling differing views regarding\\AI transformation} & \\
        \cline{2-2}
        & Intergenerational Collaboration & \\
        \cline{2-2}
        & Public sector’s unique nature & \\
        \hline
        \multirow{3}{*}{Resistance to Change} 
        & Job displacement & \multirow{3}{5.2cm}{The \textit{Resistance to Change} theme concerns common fears and conditions, which collectively create obstacles to the AI-driven Organizational Change.} \\
        \cline{2-2}
        & \makecell{Lack of deep AI knowledge\\and understanding} & \\
        \cline{2-2}
        & Lack of patience to work with AI & \\
        \hline
        \multirow{3}{*}{Skills in Era of AI} 
        & Keeping up with AI advancements & \multirow{3}{5.2cm}{That particular theme focuses on the need for skill adaptation and continuous learning together with the adoption of AI.} \\
        \cline{2-2}
        & Lack of skills due to AI & \\
        \cline{2-2}
        & Necessity to adjust skillset & \\
        \hline
        \multirow{5}{*}{Strategic Adoption of AI} 
        & \makecell{Balance automation with\\human supervision} & \multirow{5}{5.2cm}{The \textit{Strategic Adoption of AI} theme summarizes the need of integrating AI in a way that will complement the human skills.} \\
        \cline{2-2}
        & \makecell{Considering AI as a team member,\\instead of tool} & \\
        \cline{2-2}
        & Contextual Use of AI & \\
        \cline{2-2}
        & \makecell{Establish balance in using internal\\and external AI solutions} & \\
        \cline{2-2}
        & Over-reliance on AI & \\
        \cline{2-2}
        & Purpose-drive AI adoption & \\
    \hline
    \end{tabular}
    }
    \caption{identified themes and sub-themes of challenges related to the BSE-driven AI transformation}
    \label{tab:challenges}
\end{table}

\textbf{\textit{Change Management Strategy:}} Participants unanimously emphasized the importance of communicating AI-driven change, both internally and externally. P7 noted, \textit{"...if people are changing their work... they need to know what they're going to. You can't just expect people to adopt AI and be really, really efficient."}. However, conveying this change can be challenging. P3 highlighted the complexity of explaining \textit{"AI and what an AI transformation means to a large organization. It is so complex and it's not enough to say it once. You have to over communicate"}. Similarly, P4 and P5 stressed the need to convince external stakeholders of AI’s effectiveness, with P5 remarking,  \textit{"It would be a challenge to convince the world that this is actually a better solution than having a human in the loop system."}.

\textbf{\textit{Ethical Concerns:}} Ethical considerations emerged as a key topic during the interviews. P5 highlighted the challenge for smaller companies to keep up with AI-driven market demands, stating,  (\textit{"You need a huge server farm to host something that is the best of the best. I don't see any way that a smaller company can compete with that"}). Another \textit{"big blocker"} is the sensitive data organizations handle, which fosters a  \textit{"risk aversion culture"}, as noted by P7. Many interviewees emphasized the need for caution in sharing organizational data with external AI tools. P4 remarked,  \textit{"...we have to be very considerate about information that we share, of course, so they kind of said to everyone, you cannot just use the general tools available"}.

\textbf{\textit{Organizational Readiness for AI Adoption:}} Managing diverse approaches and perspectives across groups and departments emerged as a shared concern. P1 voiced frustration over project delays, stating, \textit{"I have some projects that have been paused... I know that they are working on it and they also have a lot of other things to do, maybe more prioritized, but the process is a bit unclear...”"}, while P2 talked about the importance of organizational alignment, saying, \textit{"It slows down the progress, because you have to get everyone on board... It's not until all people actually will work with it and learn how they can use this for themselves, when it will have a real effect."}. 

Another challenge will be bridging the generational gap in the coming years. P8 observed,  (\textit{"there are big differences in how people are being educated now versus 10 years ago versus 20 years ago. So we need to be prepared... creating those links... is very crucial"}). Additionally, P2 and P8 highlighted unique barriers due to the organization’s nature and being in Sweden. P2 pointed to \textit{"hurdles that the nature of public sector raises, e.g. regarding data handling,"} which must be addressed early on. P8 added, \textit{"So what we try to apply in Sweden or what we try to apply globally does not always land the same way."}. 

Finally, P2 and P8 discussed challenges in cross-functional collaboration. For example, P2 noted,  \textit{"It's really hard to make the lawyers to understand what it is. So it's hard for us... it's more about that the knowledge is on different levels and that creates conflicts and friction."}

\textbf{\textit{Resistance to Change:}} Resistance to change emerged as a major issue discussed by all participants. Job security concerns contributed to this resistance, with P7 noting,  \textit{"I mean, there's a fear of the unknown. And, you know, what is this going to be like? Will I have a job?"}. However, resistance was not solely driven by fears of job displacement. Limited AI knowledge also played a role; as P4 explained, \textit{"...there's some new things to learn that are unfamiliar to me... I've been working for 10 years so I never had the opportunity to study it and learn about it like those in college now."}). Additionally, a lack of patience appeared to sometimes fuel resistance. P2 observed, \textit{"People really think that all the IT products are black or white... when they don't get the right result, they just consider this product as obsolete or useless."}

\textbf{\textit{Skills in Era of AI:}} The rapid evolution of skills and roles driven by AI change, poses significant challenges for both employees and organizations. Participants remarked on the unprecedented pace of change; P7, referencing a former Google employee, noted that staying current requires \textit{"three hours per day to keep up to date with the changes"}. P5 emphasized the need to adapt skill sets to meet AI demands, such as  \textit{"by learning how to write prompts to an AI"}. However, P6 expressed concern about future employees becoming overly reliant on AI, maybe already in college, cautioning, \textit{"...being over-reliant on this GPT or AI makes you personally inefficient. So let's say without GPT or AI, your skills would be zero in that case in future"}.

\textbf{\textit{Strategic Adoption of AI:}} The strategic integration of AI presents challenges, particularly in maintaining a balanced approach that avoids over-reliance and hindering human efforts. P3, for instance, suggested, (\textit{"...sometimes we have to take a step back and say do we need AI or can we just do this with regular automation?"}). Likewise, P10 explained the provocation of finding the \textit{"balance between automation and human presence, as human knowledge will be more essential in the future"}. P2 highlighted the importance of understanding data biases and using AI tools appropriately, noting, \textit{"I think you have to be aware of the biases in data. Based on that you have to think in which areas the product can be good and in which areas it won't be good"}. Similarly, P8 discussed the importance of establishing a clearer balance and guidelines on using internal versus external AI solutions to \textit{"increase people's confidence at how they can use each solution"}. An interesting perspective was shared by P5 and P8 regarding the feasibility of integrating AI as a group member, so that it will complement and enhance the quality of the human efforts. More specifically, P5 questioned; \textit{"AI is a tool that everyone uses in their screens; would we be able to integrate it as an actual part of the group?"}. Finally, P9 underscored the \textit{necessity of having identified a particular problem, before starting adopting AI.}. Otherwise, it constitutes a \textit{pointless ''Transformation''}, as they stated.

\subsection{RQ3: How the different roles experience AI Transformation}

After the thematic analysis, narrative analysis was applied to answer the third research question of the study on the potentially differing experiences of different roles. The analysis of the transcripts revealed that, while each role shares similarities in their experience of the AI transformation journey, they each focus on distinct concepts and challenges. A summary of the findings is presented in \ref{tab:narrative_analysis}. As previously explained, narratives were categorized as progressive, stable, and regressive, indicating a positive, negligible, and negative impact on individuals' roles and experiences, respectively.

\begin{table}[h!]
    \centering
    \scalebox{0.65}{
    \begin{tabular}{|c|c|c|}
       \hline
        \textbf{Role} & \textbf{Interpretation} & \textbf{Summary} \\ 
        \hline
        \multirow{4}{*}{Developer} 
        & \multirow{4}{5cm}{Progressive narrative in general, accompanied with a reserved optimism due to expressed job security concerns.} & \multirow{4}{5.2cm}{Early-adopters from their nature, Developers embrace the AI adoption. They identify the need for skills adaptation, but they, also, feature the job security concerns.} \\
        & & \\
        & & \\
        & & \\
        \hline
        \multirow{4}{*}{HR Roles} 
        & \multirow{4}{5cm}{Stable narrative, as the AI-driven Organizational Change is at an early stage.} & \multirow{4}{5.2cm}{The AI integration has not significantly affected their current workflow. Overall, mixed feelings were observed regarding the impact of AI and the anticipation of change.} \\
        & & \\
        & & \\
        & & \\
        \hline
        \multirow{4}{*}{IT Section Manager} 
        & \multirow{4}{5cm}{Progressive narrative, which conveys high optimism and AI acceptance.} & \multirow{4}{5.2cm}{AI has already influenced the workflow. They view AI as an opportunity, identifying the challenges that have already been brought in the management level.} \\
        & & \\
        & & \\
        & & \\
        \hline
        \multirow{4}{*}{\makecell{Roles created due to\\AI Transformation}} 
        & \multirow{4}{5cm}{Progressive narratives, as their positions are a result of the AI Transformation.} & \multirow{4}{5.2cm}{Being part of the AI strategy, they perceive AI as an enabler for further development and progress. However, they deal with many barriers and existing social issues.} \\
        & & \\
        & & \\
        & & \\
        \hline
        \multirow{4}{*}{System Engineers} 
        & \multirow{4}{5cm}{Stable narratives, as the AI Transformation journey is at the very beginning for their organization.} & \multirow{4}{5.2cm}{Their role is unchanged during the early phases of the Transformation. They acknowledge the potential, having a more reserved approach though.} \\
        & & \\
        & & \\
        & & \\
    \hline
    \end{tabular}
    }
    \caption{narrative analysis results of how different roles experience AI Organizational Change}
    \label{tab:narrative_analysis}
\end{table}

As shown in Table \ref{tab:narrative_analysis}, none of the roles included in this research exhibited a regressive narrative. 

\textbf{\textit{Progressive narratives:}}  The Developers (P6, P10), IT Section Manager (P2), and roles specifically created for the AI transformation (P1, P3) described primarily positive impacts of AI on their workflows. A shared pattern among these roles is a generally positive attitude toward AI’s growth, despite challenges along the way. For example, P3 stressed \textit{"I think it will remove the monotonous work, will hopefully elevate the workday. I think it will be a lot better, more productive, more fulfilling."}. Despite the progressive narratives, participants shared challenges that they are communicated within their working environment. P6 voiced \textit{"concerns about job security"}, a sentiment shared by other developers involved in manual tasks, emphasizing on the need of working in your skills. Otherwise, individuals will end up \textit{"having zero skills without AI and GPT"}. Similarly, P10 shared the \textit{"importance of having a clear strategy and planning in AI integration"}, acknowledging though the \textit{"confusion and uncertainty"} due to the uncharted nature of AI.

P1, P2, and P3 focused more on provocations in communicating the change and handling differing strategies and priorities without causing delays in projects. P1 explained \textit{"They are falling behind, and there's a lot of discussion in the organization on how can get them included,"}, while P2 noted \textit{"You have to get everyone on board to make really good things with AI."}. These narratives largely focus on strategies to overcome challenges and address related social issues, such as emphasizing AI training and skills adaptation to help organizations alleviate concerns about the future job market.
    
\textbf{\textit{Stable narratives:}} System engineers (P4, P5) and HR roles (P7-P9) shared stable narratives, indicating that AI has not yet significantly impacted their current roles, as their organizations are still in the early stages of AI adoption. P4 and P5 expressed optimism about AI’s potential and impact on organizations workflow. P5 highlighted \textit{"...but with a better AI, maybe it could also handle an unforeseen event like that, increasing safety and quality."}. However, both System Engineers had a more reserved approach toward its integration, likely influenced by their \textit{"organization’s cautious stance on adopting new technologies"}, as they noted.

HR roles, namely P7, P8, and P9, expressed divergent perspectives towards AI integration. P7 and P8 were more motivated to engage with AI, often experimenting with it outside of work. P8 shared their impression that \textit{"AI can become a partner to do work with."}, while both P7 and P8 discussed their anticipation for the impact of AI in HR repetitive tasks to \textit{"free up their work"}, as P7 voiced. Their narratives highlighted the constructive role of AI in fostering creativity and other innovations. P9 held a more reserved approach towards AI, connecting to their overall conservatism towards adopting new technologies. Consequently, their narrative communicated concerns for the \textit{"AI hype"} and the general tendency to integrate AI without having identified particular needs. However, they expressed their reserved optimism that AI tools will be salutary for their organization.

\section{Discussion}

This study investigated how artificial intelligence drives software organizational change from a behavioral software engineering (BSE) perspective.
Our analysis of the semi-structured interviews highlight the importance of human factors in the AI transition, underscoring that AI transformation is not solely technical but deeply intertwined with human behaviors and attitudes. Thematic analysis revealed twelve BSE concepts and several behavioral sub-themes across three analytical levels, along with six primary themes on the challenges organizations face in adopting AI.

The participating organizations are in early stages of AI transformation, which concentrated our findings at the individual level. We hypothesize that additional group and organizational level findings can be found by studying organizations that have been working with and changing due to AI for longer.

Personality and personal variations emerged as important factors. Employees with higher openness to change showed more positive emotions and attitudes toward AI, in line with earlier findings about personality and attitude correlations~\cite{feldt2010personalitiesAndAttitudes}. For example, P7's initial excitement about AI tools led them to embrace change and experiment with AI to enhance their creativity. In contrast, more skeptical participants like P5 worried about AI's impact on future employment. This reserved stance is, likely, linked to the culture of their organizations, as P4, P5, and P9 who all demonstrated more cautious approaches, have been part of organizations with more conservative strategies for several years.
While prior research has highlighted novelty and personal growth as motivations to use AI~\cite{skjuve2024ChatGPTMotivation}, we found that personal traits like curiosity and a desire to automate repetitive tasks also seemed to drive AI integration.

Participants frequently cited confusion and misconceptions around AI, which we categorized as cognitive issues—a topic seldom addressed in BSE literature. Given AI's continued growth and diversification, it is crucial for companies to actively support employee education and foster a nuanced understanding of AI. Such an approach helps employees engage with AI on a concrete level, reducing the risk of abstract discussions that may fuel apprehension or resistance.

Emotional responses to change clearly varied widely among interviewees, from excitement and curiosity to frustration and apprehension, showing the complexity of human adaptation to AI and underscoring emotions as a critical factor in the adoption of new technologies.

At the group level, participants indicate that integrating AI has influenced group dynamics,  emphasizing Agile methods over traditional approaches. This shift requires greater collaboration, cross-functional teamwork, but potentially also short-term goal setting, due to AI’s multifaceted impacts across technical, legal, and behavioral domains. Interestingly, politics surfaced as a significant theme at the group level, with differences in roles, mindsets, and strategies leading to project delays. Politics is not extensively discussed in the Group Level, so future research could provide a deeper analysis of how internal politics within teams impact the successful adoption of AI and of similar projects.

At the organizational level, three main themes emerged: Organizational Culture, Organizational Adaptability, and Politics. Participants reported varied organizational values and practices in AI integration, heavily influenced by each organization’s culture and maturity in handling change. A common thread was the need for a culture of experimentation and knowledge-sharing around AI, facilitated by training and workshops to address cognitive gaps. Organizational Adaptability, a relatively new concept within BSE and the work psychology literature, surfaced as essential, with strategic planning needed to align AI advancements with organizational goals and mitigate budget misalignments and employee frustration. Rapid AI development also implies a need for flexible, iterative planning.

Politics at the organizational level involved both internal and external pressures for AI adoption, with departments sometimes competing over similar projects, leading to resource inefficiencies and fragmentation. P3, for instance, described instances where multiple departments independently pursued the same initiative, emphasizing the need for coordinated management of political dynamics to prevent such overlap.

While the study yielded a rich picture at the individual level, this should not detract from the insights gained at group and organizational levels. The organizations’ early stage of AI transformation suggests that further research is needed to understand the evolving impact of AI on organizations as they mature in this area. A long-term perspective will help reveal the broader organizational effects of AI.

This study also identified challenges specific to AI transformation in software organizations, resonating with change management factors identified by Errida and Lotfi \cite{Errida2021DescriptiveOrgModel}, including continuous communication, establishing a clear vision, resistance management, and leadership initiatives. These elements align closely with the concerns of study participants, with leadership and resistance management cited as particularly relevant in addressing resistance to change. Future research could involve more empirical data to challenge and enrich the existing theoretical change management models.

Beyond challenges, the study explored how various roles experience AI transformation. None of the participants reported a regressive narrative characterized by negative experiences like stress or job insecurity.
Rather, participants reported progressive or stable narratives without strong negative sentiments. Nonetheless, ethical and human concerns, such as privacy, remain central to participants’ reflections, highlighting the importance of considering these dimensions in future organizational strategies.

This research offers insights for both practitioners and researchers. For practitioners, the identified BSE concepts and associated challenges highlight the importance of human factors in successful AI integration. A deeper understanding of these human dimensions can help better support and create environments for smoother transitions to AI-driven workflows. For instance, tailored programs based on diverse attitudes toward change, offering advanced workshops for employees with high curiosity and openness and more foundational training for those more reserved to build trust and confidence. 

While adopting an Agile culture, encouraging experimentation and innovation are, arguably, already common in software organizations, several of our findings also have implications for other domains. Considering the ethical dimension of the AI integration, such as the data privacy, is essential in all fields. Furthermore, the necessity for AI training and skill adaptation is related to all fields that are affected by AI transition, especially in non-technical domains that employees might lack fundamental technical knowledge. Training and continuous communication are key enablers for AI adoption across fields to deal with resistance to change.

For researchers, this study contributes to the understanding of human and cooperative aspects on the AI transformation in software organizations. The twelve BSE concepts and six identified challenges provide a foundation for further investigation, inviting deeper exploration of themes like Motivation and Politics at the group level. Future research can also work toward establishing best practices for AI-driven organizational change, contributing to the development of theoretical models in this evolving field.

\section{Conclusions}

AI is increasingly adopted and already affecting organizational structures and procedures~\cite{subramanian_emergent_2017}. Software development and software-dependent organizations, in particular, are changing. However, behavioral and psychological research in this area remains limited \cite{peretz_andersson_empirical_2022}. To address this gap, we conducted a qualitative study, by collecting and analyzing ten semi-structured interviews with industry practitioners, to better understand AI-driven software organizational change. Our analysis was done from a Behavioral Software Engineering (BSE) perspective and the findings structured based on core concepts and levels of analysis from the BSE field. This exploratory study employed a purposeful sampling strategy to capture diverse insights from participants across varied roles and departments in four different organizations.

Through thematic analysis of the interview transcripts, we identified several sub-themes under twelve core BSE concepts, categorized into three levels of analysis: \textit{Attitudes, Cognitive, Creativity, Emotions, Group Dynamics, Motivation, Organizational Culture, Organizational Readiness, Politics (both at the Group and Organizational level), Personality, and Stress}. While most of the findings focus on the individual level, it is important to highlight that the complexities at the group and organizational levels likely cannot be fully captured yet. As organizations continue their AI transformations, future studies are necessary to establish a fuller and more mature picture also at these levels of analysis. Additionally, six challenges was found in relation to the identified BSE themes: \textit{Communication Strategy, Ethical Concerns, Organizational Readiness for AI Adoption, Resistance to change, Skills in Era of AI, and Strategic Adoption of AI}. We discuss how these themes intersect with existing literature and relate to established change management frameworks.

The narrative analysis revealed that none of the involved roles use a regressive narrative of the AI transformation. However, both progressive and stable groups of narratives expressed specific concerns that organizations must carefully consider, particularly regarding human and ethical aspects. 

Finally, the study discusses future research directions on further exploring particular BSE concepts within the levels of analysis, and on unraveling the good practices for a BSE-driven AI Organizational Change, by challenging and enriching existing theoretical frameworks.

\section*{Acknowledgment}

The authors want to thank all the nine participants for devoting their time to take part in the research and the individuals who helped approaching them, providing valuable help throughout the process.

\bibliographystyle{IEEEtran}
\bibliography{IEEEabrv, references}

\end{document}